\documentclass[pre,
               showpacs,
               superscriptaddress,
               longbibliography
               ]{revtex4-1}

\usepackage{graphicx}% Include figure files
\usepackage{dcolumn}% Align table columns on decimal point
\usepackage{bm}% bold math
\usepackage{amsmath}
\usepackage{amssymb}
\usepackage{color}

\begin{document}

\title{``Eppur \emph{non} si muove'': the effect of roll number on the statistics of turbulent Taylor-Couette flow}

\author{Rodolfo Ostilla-M\'{o}nico}\email{rostillamonico@g.harvard.edu}
\affiliation{Physics of Fluids Group, Faculty of Science and Technology, MESA+ Research
             Institute, and J. M. Burgers Centre for Fluid Dynamics,
             University of Twente, PO Box 217, 7500 AE Enschede, The Netherlands.}
\affiliation{School of Engineering and Applied Sciences, Harvard University, Cambridge, MA, USA.}

\author{Detlef Lohse}
\affiliation{Physics of Fluids Group, Faculty of Science and Technology, MESA+ Research
             Institute, and J. M. Burgers Centre for Fluid Dynamics,
             University of Twente, PO Box 217, 7500 AE Enschede, The Netherlands.}

\author{Roberto Verzicco}
\affiliation{Dipartimento di Ingegneria Industriale, University of Rome ``Tor Vergata", Via del Politecnico 1, Roma 00133, Italy}
\affiliation{Physics of Fluids Group, Faculty of Science and Technology, MESA+ Research
             Institute, and J. M. Burgers Centre for Fluid Dynamics,
             University of Twente, PO Box 217, 7500 AE Enschede, The Netherlands.}
             
\date{\today}

\begin{abstract}

A series of direct numerical simulations in large computational domains has been performed in order to probe the spatial feature robustness  of the Taylor rolls in turbulent Taylor-Couette (TC) flow. The latter is the flow between two coaxial independently rotating cylinders of radius $r_i$ and $r_o$, respectively.  Large axial aspect ratios $\Gamma = 7$-$8$ (with $\Gamma = L/(r_o-r_i)$, and $L$ the axial length of the domain) and a simulation with $\Gamma=14$ were used in order to allow the system to select the most unstable wavenumber and to possibly develop multiple states. The radius ratio was taken as $\eta=r_i/r_o=0.909$, the inner cylinder Reynolds number was fixed to $Re_i=3.4\cdot10^4$, and the outer cylinder was kept stationary, resulting in a frictional Reynolds number of $Re_\tau\approx500$, except for the $\Gamma=14$ simulation where $Re_i=1.5\cdot10^4$ and $Re_\tau\approx240$. The large-scale rolls were found to remain axially pinned for all simulations. Depending on the initial conditions, stable solutions with different number of rolls $n_r$ and roll wavelength $\lambda_z$ were found for $\Gamma=7$. The effect of $\lambda_z$ and $n_r$ on the statistics was quantified. The torque and mean flow statistics were found to be independent of both $\lambda_z$ and $n_r$, while the velocity fluctuations and energy spectra showed some box-size dependence. Finally, the axial velocity spectra was found to have a very sharp drop off for wavelengths larger than $\lambda_z$, while for the small wavelengths they collapse.
\end{abstract}


\maketitle

%\section{Introduction}

Taylor-Couette (TC) flow, the flow between two independently rotating co-axial cylinders is one of the paradigmatical systems in fluid mechanics, both due to its high simplicity and its applications in technology and Nature. While the low Reynolds number regime of TC flow has been studied in great detail for decades \cite{far14}, the large Reynolds number regime remained relatively unexplored until the last few years \cite{gro16}. Only recently enough computational power has become available that fully resolved simulations reaching the so-called ``ultimate'' regime of Taylor-Couette flow, where both boundary layers and bulk are turbulent, have become a possibility \cite{ost14}. 

In absence of viscosity, TC flow is linearly unstable if $|r^2_i\omega_i|>|r^2_o\omega_o |$, where $r_i$ and $r_o$ are the inner and outer cylinder radia, respectively, and $\omega_i$ and $\omega_o$ are the inner and outer cylinder angular velocities. Due to this instability, a series of transitions take place when the inner cylinder is rotated with increasing speed. For very small driving, the flow is purely azimuthal. Once the driving is large enough to overcome the viscous damping, this purely azimuthal flow becomes centrifugally unstable and stationary large-scale structures fill the entire gap, effectively redistributing angular momentum. These structures are called Taylor rolls after the seminal work by Taylor \cite{tay23}. Further increasing the driving causes the onset of time-dependence, and the Taylor rolls to transition to wavy Taylor rolls first, then to modulated Taylor rolls, and finally to turbulent Taylor rolls \cite{and86}. At the highest Reynolds numbers achieved in both simulations, i.e.~$Re\sim\mathcal{O}(10^5$)  and experiments i.e.~$Re\sim\mathcal{O}(10^6)$, an axially stationary signature of these rolls can be observed \cite{hui14,ost16}. This signature is only present in certain regions of the high Reynolds number TC flow parameter space, mainly depending on the radius ratio of the system $\eta=r_i/r_o$ and the rotation ratio $\mu=\omega_o/\omega_i$  \cite{ost14d}. 

Experimental realizations of TC flow necessarily have end-caps at the top and bottom of the systems, which may be fixing the position of the rolls in the axial direction. The first photographs of experimentally pinned Taylor rolls were provided by Coles \cite{col65}, at Reynolds numbers of  $Re\sim\mathcal{O}(10^4)$. Further studies by Benjamin \& Mullin \cite{ben82}, Andereck \emph{et al.}, \cite{and86}, Lathrop \emph{et al.} \cite{lat92a}, Martinez-Arias \emph{et al.} \cite{mar14} and Huisman \emph{et al.} \cite{hui14} have repeatedly shown, in several experiments up to  $Re\sim\mathcal{O}(10^6)$ that the rolls are pinned, and have also shown the multiplicity of roll-states and the crucial role of the initial conditions and hysteresis in determining the aspect ratio of the rolls. A detailed study of roll-size hysteresis at $Re\sim\mathcal{O}(10^6)$ was performed by van der Veen \emph{et al.} \cite{vee16}.  Simulations use periodic boundary conditions, which a priori should not fix the position of these rolls. The puzzling axial pinning of the Taylor rolls, observed in Refs. \cite{ost14d,ost16} and the resulting lack of statistical axial homogeniety of the axially-periodic direct numerical simulations (DNS) was speculated to be caused by an insufficient axial extent of the domain \cite{jobp15}. ``Small'' computational boxes have been used in TC flow to be able to perform the high Reynolds number simulations by both Brauckmann \emph{et al.} \cite{bra13,bra13b,bra15}, and by Ostilla-M\'onico \emph{et al.} \cite{ost14,ost14d,ost16}. In these simulations, the aspect ratio $\Gamma=L_z/(r_o-r_i)$, where $L_z$ is the axial periodicity length, was limited to $\Gamma\approx2$, enough to fit a single roll pair. In addition, a rotational symmetry of order $n_{sym}$ was imposed to reduce the azimuthal extent of the domain at the mid-gap $L_x$ to $L_x/d\approx \pi$. To assess the validity of these computational boxes, a systematic study was conducted by Ostilla-M\'onico \emph{et al.}~\cite{ost15}, who found that these small boxes were sufficient to produce box-independent statistics for the torque and mean velocity profiles for pure inner cylinder rotation at $Re_i=10^5$, where $Re_i$ is the inner cylinder Reynolds number is defined as $Re_i=(r_o-r_i)\omega_ir_i/\nu$, and $\nu$ is the kinematic viscosity of the fluid. However, the velocity fluctuations were found to be box-dependent, even for the largest boxes considered. This trend was in agreement to what was observed in DNS of channel flow, i.e.~the pressure driven flow between two parallel plates, by Lozano-Dur\'an and Jim\'enez~\cite{loz14}.

A second raised issue was that in Ref.~\cite{ost15} only cases with a single roll pair were considered. The aspect ratio $\Gamma$ was varied between $2$ and $4$, and in the simulations, a single, axially stationary roll pair was observed, whose wavelength $\lambda_z$ was found to grow with increasing $\Gamma$ up to $\lambda_z=4$. For even larger $\Gamma$, one could expect more than a single roll pair to form, and even to have multiple ``states'', i.e.\ different number of turbulent roll pairs depending on the initial conditions, which could affect the statistics dramatically \cite{hui14}.

In this manuscript we conduct a series of DNS of TC flow using the incompressible Navier-Stokes equations in computational boxes with very large axial and azimuthal extents, to answer the issues previously raised. These boxes can fit more than a turbulent Taylor roll pair, and thus can be used to assess the effect of the roll number on high-order statistics. These simulations were performed using an energy-conserving second-order finite difference code \cite{ver96,poe15}, with fractional time-stepping. The radius ratio was fixed to $\eta=0.909$, in the parameter space region where small boxes showed Taylor rolls to be strong and axially pinned. A rotational symmetry $n_{sym}=5$ was imposed, meaning that only a fifth of the cylinder was simulated and periodic boundary conditions were used in the azimuthal direction. This is a smaller $n_{sum}$ than previously used for this $\eta$ ($n_{sym}=20$), so the domain was four times larger in the azimuthal direction. In the axial direction, $\Gamma$ was chosen to be either $\Gamma=7$ or $\Gamma=8$. This results in a streamwise extent of the box at mid-gap of $8.4\pi$ half-gaps and an axial extent of the box of $14$ ($4.5\pi$) or $16$ ($5.1\pi$) half-gaps, comparable to the large boxes run in plane Couette (PC) flow simulations \cite{avs14,pir14}. PC flow is the flow between two parallel and independently moving plates. Rotating PC flow, where the two plates can also rotate about an axis parallel to them is the limiting case of TC flow when $\eta\to 1$. Rotating PC flow has two control parameters, the shear Reynolds number and the Rotation/Rossby number, which are equivalent to the shear Reynolds number and Rossby number for TC flow as defined in Ref.~\cite{dub05}.  Unlike TC flow, PC flow does require large computational boxes as the decorrelation lengths are much longer \cite{tsu06}.

In these simulations, the inner cylinder Reynolds number was set to $Re_i=3.4\cdot10^4$, while the outer cylinder was kept stationary. This resulted in an inner cylinder frictional Reynolds number $Re_\tau=u_{\tau,i}(r_o-r_i)/(2\nu)\approx 500$, where $u_{\tau,i}$ is the inner frictional velocity $u_{\tau,i} = (\tau_i/\rho)^{1/2}$, $\tau_i$ the shear stress at the inner cylinder and $\rho$ the fluid density. The outer cylinder frictional velocity (Reynolds number) is given by $u_{\tau,o}=\eta u_{\tau,i}$ ($Re_{\tau,o}=\eta Re_{\tau,i})$. For convenience, we also define the non-dimensional distance from the wall $\tilde{r}=(r-r_i)/(r_o-r_i)$, the non-dimensional axial coordinate $\tilde{z}=z/(r_o-r_i)$, and the non-dimensional azimuthal coordinate at the mid-gap $\tilde{x}=\frac{1}{2}(r_i+r_o)\theta/(r_o-r_i)$. 

The simulations were ran in a rotating frame of reference similar to the one proposed by Dubrulle \emph{et al.}~\cite{dub05}, such that the velocity at both cylinders was equal to a half of the characteristic velocity, and of opposite sign, and thus the mean velocity was equal to zero at the mid-gap to reduce as far as possible the dispersion errors in the spectra due to the use of finite differences and allows for larger time steps for the same $\Delta t^+$ \cite{ber13}. We note that the Reynolds numbers simulated here are a factor three smaller than the one considered for the previous box-size comparison in Ref.~\cite{ost15}, i.e.~$Re_\tau\approx 1400$ and $Re_i=10^5$, but this is necessary to keep the computational costs manageable with large computational domains. With this Reynolds number, the largest grid resolution used was $N_\theta\times N_r \times N_z=1536\times512\times2926$ for the $\Gamma=8$ box, i.e.~over two billion points in the largest simulation. Full details of the numerical resolutions used are in Table~\ref{tbl:reso}. Points were clustered in the radial direction using a clipped Chebychev distribution, and homogeneously distributed in the other two directions, which resulted in a resolutions in inner cylinder wall units of $r_i\Delta \theta^+=8.6$, $\Delta r^+ \in (0.4,2.9)$ and $\Delta z^+=2.7$, where wall units are defined using $u_\tau$ and $\delta_\nu=\nu/u_\tau$.  The timestep of the simulations was taken so that $\Delta t^+=0.4$, and the simulations were ran for over $100$ large eddy turnover times based on $d/(r_i\omega_i)$, equivalent to at least $6$ turnover times based on the frictional velocity and the half-gap, i.e.~$d/(2u_\tau)$. This resulted in a wall-time of three weeks on $480$ cores. The statistical convergence of the solution can be estimated by noting that the angular velocity current $J^\omega=r^3(\langle u_r \omega \rangle_{\theta,t,z} - \nu \partial_r \langle \omega \rangle_{\theta,t,z})$ is radially constant to within $1\%$, where the $\langle ... \rangle_{x_i}$ operator indicates averaging with respect to the independent variable $x_i$. 
 
\begin{table}
  \begin{center} 
  \def~{\hphantom{0}}
  \begin{tabular}{|c|c|c|c|c|c|c|c|c|c|c|}
  \hline
  Case & $\Gamma$ & $\lambda_z$ & $n_{sym}$ & $N_\theta$ & $N_r$ & $N_z$ & $Nu_\omega$ & $Re_{\tau,i}$ & $t_{stat}U_i/d$ & Colour of lines \\ 
  \hline
  G2R1 & $2.33$ & $2.33$ & $20$ &  $384$ & $512$ & $768$ & $24.3$ & $490$ & $233$ &Blue solid \\
  G7R3 & $7$ & $2.33$ & $5$ & $1536$ & $512$ & $2560$ & $24.3$ & $490$ & $78$ &Orangish solid \\
  G7R2 & $7$ & $3.5$ & $5$ & $1536$ & $512$ & $2976$ & $26.3$ & $510$ & $121$ & Ocre dashed\\
  G8R3 & $8$ & $2.66$ & $5$ & $1536$ & $512$ & $2976$ & $24.9$ & $500$ & $104$ & Purple dash-dot \\ 
  G14R7 & $14$ & $2$ & $2$ & $1536$ & $256$ & $2560$ & $13.5$ & $240$ & $487$ & None \\ 
    \hline
 \end{tabular}
 \caption{Details of the numerical simulations.  The first column is the case name, the second column is the aspect ratio $\Gamma$, the third column is the roll wavelength $\lambda_z=\Gamma/n_r$, where $n_r\in \mathbb{N}$ is the number of rolls. The fourth column is the imposed rotational symmetry. The fifth to seventh columns represent the amount of points in the azimuthal, radial and axial directions. The eighth column represents the non-dimensional torque $Nu_\omega$. The ninth column is the frictional Reynolds number at the inner cylinder.  The second to last column shows the averaging time for statistics. The last column refers to the line colors in the  Fig.\ \ref{fig:q1wall} to \ref{fig:spectraslab9}. }
 \label{tbl:reso}
\end{center}
\end{table}

\begin{figure}
  \centering
    \includegraphics[width=0.80\textwidth]{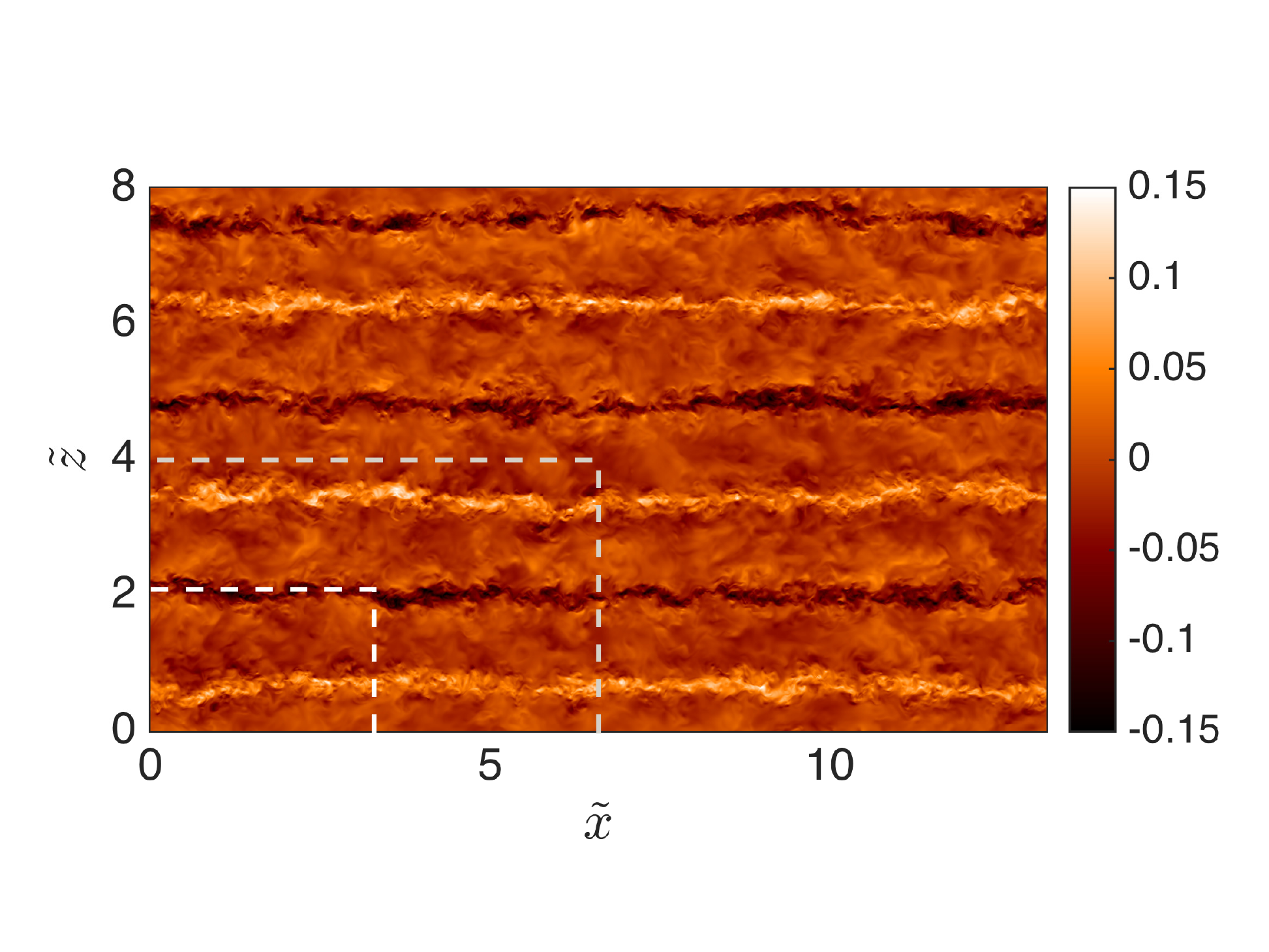}
  \caption{ Instantaneous angular velocity at the mid-gap for the $\Gamma=8$ simulation. $\tilde{z}=z/d$ is the axial direction and $\tilde{x}=r_a\theta d$ the rescaled azimuthal direction with $r_a$ the mid-gap radius. The signature of the Taylor rolls can be seen as a large, stripy pattern, which appears to be fixed in the axial position for the entire azimuthal extent. For comparison, the (smaller) white red line indicates the typical size of a small TC flow computational box of \cite{ost16}. The (larger) light grey dashed line indicates the largest box simulated in Ref.~\cite{ost15}.  }
\label{fig:box_measurement}
\end{figure}

First, two simulations, one for $\Gamma=7$ and one for $\Gamma=8$, were started from initial conditions consisting of a quiescent fluid with some added white noise in the velocity fields. During a transient which lasted around $200 d/(r_i\omega_i)$ turnover times, large-scale axially-stationary rolls were seen to form, two pairs in the case of $\Gamma=7$ (this would be the G7R2 simulation) and three pairs in the case of $\Gamma=8$ (G8R3), resulting in roll wavelengths of $\lambda_z=3.5$ and $\lambda_z=2.66$ respectively. Figure \ref{fig:box_measurement} shows a pseudo color plot of the azimuthal velocity at the mid-gap for the largest box simulated, and a sketch of where the ``small'' boxes normally used for TC flow DNS, and of the largest box of Ref.~\cite{ost15} for comparison purposes. 

After this, the statistically stationary flow field from the G8R3 case was rescaled to fit into a $\Gamma=7$ box, and used as initial condition for a third case, G7R3. This case was advanced in time for around $300 d/(r_i\omega_i)$, around $200$ turnover times to overcome the transient and $100$ more to take statistics. The resulting statistically stationary flow field from this third simulation had a different roll state- i.e.~three roll pairs, with wavelength $\lambda_z=2.33$, different from the two roll pair $\lambda_z=3.5$ case arising from the white noise initial conditions. We note that this axial re-scaling, while convenient for generating the desired roll states, does not always work. This method cannot generate unphysical wavelengths since if the rolls are stretched or contracted beyond physical solutions, they will merge or break up. Finally, to asses the effect of the amount of rolls on the flow statistics, a fourth simulation (G2R1) with $\Gamma=2.33$ and a single roll pair of $\lambda_z=2.33$ was performed to compare against the G7R3 case with 3-roll pairs of $\lambda_z=2.33$. 

To ensure that the rolls were indeed fixed even for larger computational boxes, one more case denoted by G14R7 was run for $Re_i=1.57\cdot10^4$. The box parameters were $\Gamma=14$ and $n_{sym}=2$, resulting in a computational box of $21\pi \times 2 \times 9\pi$ half-gaps. The resulting inner cylinder frictional Reynolds number was $Re_\tau \approx 240$. As $Re_\tau$ was smaller, this allowed for coarser grids and larger time steps while still maintaining the accuracy of the simulations. The simulations were started from white noise initial conditions, and a state with seven rolls formed. This case was run for an even longer time to collect statistics: 487 time-units based on the large-eddy turnover time, or 29 time-units based on the frictional velocity and the half-gap.

Table \ref{tbl:reso} provides a summary of the simulations ran, and the resulting frictional Reynolds numbers and non-dimensional angular velocity current (torque) pseudo-Nusselt number $Nu_\omega = J^\omega/J^\omega_{pa}$ \cite{eck07b}, where $J^\omega_{pa}$ is the angular velocity current for the purely azimuthal case. A very weak variation of the frictional Reynolds number can be seen, which is consistent with the weak dependence of the torque on the roll wavelength for the Reynolds number considered here \cite{ost14d}. The $Nu_\omega$ for the single and three roll pair cases at the same $\lambda_z$ coincides within statistical convergence, consistent with the low Reynolds number result of Ref.~\cite{bra13}. As a consequence, the resulting $Re_\tau$, which scales as $Re_\tau\sim\sqrt{Re_iNu_\omega}$ is also independent of the number of rolls. The small dependence of $Nu_\omega$ on $\lambda_z$ vanishes for larger Reynolds number so box-independent values for the torque can be obtained for $\Gamma=2$ \cite{ost14d,ost15}.

\begin{figure}
  \centering
    \includegraphics[height=9cm,trim={7cm 0 7cm 0},clip]{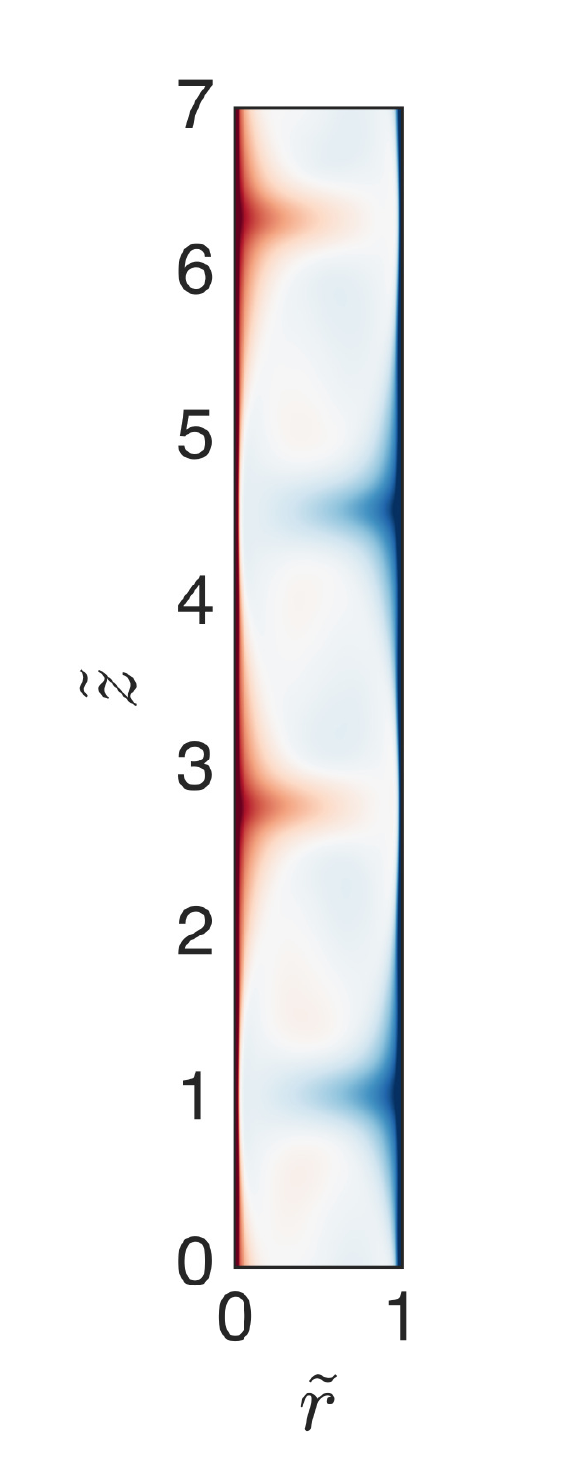}
     \includegraphics[height=9cm,trim={6cm 0 7cm 0},clip]{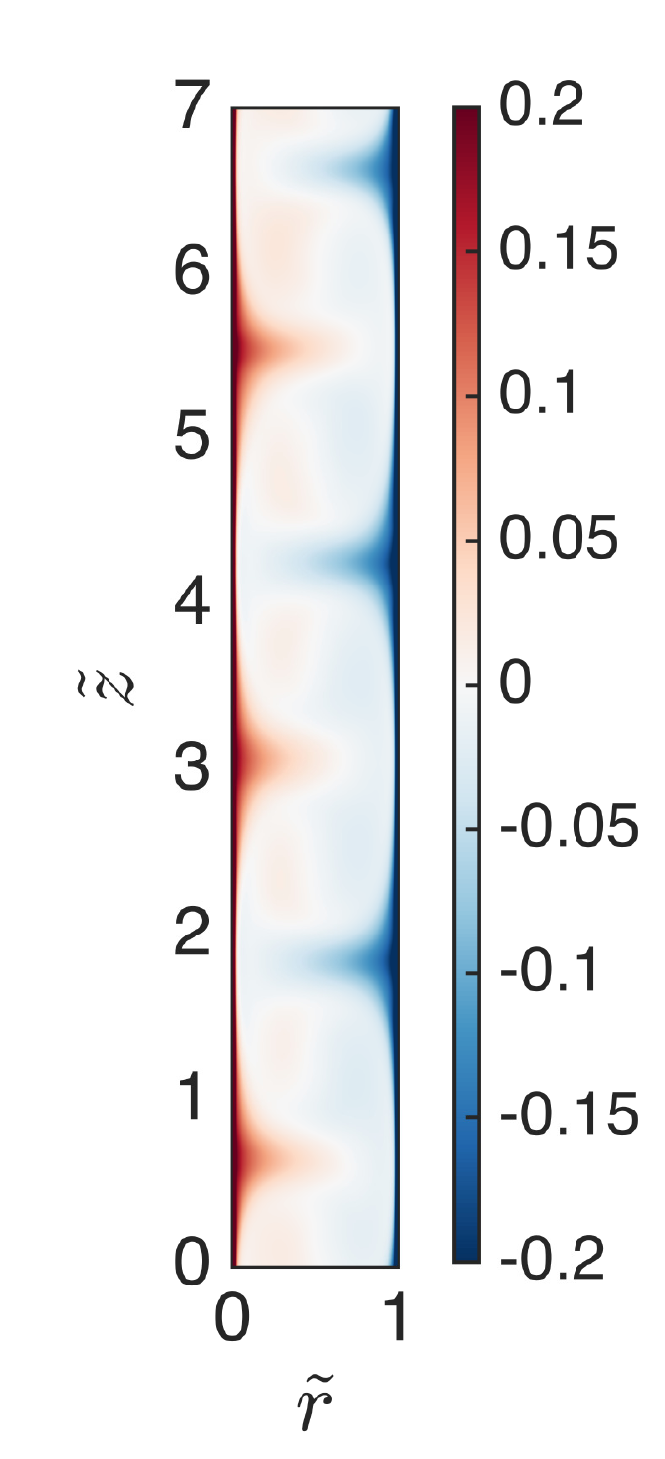}
  \caption{ Azimuthally- and temporally averaged azimuthal velocity for the G7R2 and G7R3 cases, starting from a random velocity field (left, G7R2) or from a rescaled velocity field with three rolls (right, G7R3 case). From the characteristic signature of the rolls, the different roll wavelength can be appreciated. Both cases have been run for about $100$ large eddy turnover times based on the large-eddy turnover time, and appear to be stable. }
\label{fig:q1meG7}
\end{figure}

Figure \ref{fig:q1meG7} shows a comparison of the azimuthally- and temporally averaged azimuthal velocity for both $\Gamma=7$ cases. As noted previously, \emph{``eppur non si muove''}, even for these simulations with very large computational boxes, the large-scale rolls are still axially fixed. We note that the roll wavelengths which result from the white noise initial conditions in large $\Gamma$ simulations are those of ``tall'' rolls for the higher Reynolds number case, and ``square'' for the lower Reynolds number case. Indeed, it seems to be the case that for increasing Reynolds numbers the characteristic roll-wavelength, and thus the axial decorrelation length \emph{increases}. This effect was already noted previously in TC flow by several authors \cite{mar14,ost14d} who observed ``tall'' rolls only for high Reynolds numbers, and ``wide'' rolls only at low Reynolds numbers. Indeed, experimentally Huisman \emph{et al.} \cite{hui14} observed rolls with wavelengths up to $\lambda_z=3.9$ at $Re\sim\mathcal{O}(10^6)$ with $\Gamma=11.7$, sufficiently large not to substantially constrain the roll wavelength. Theoretical results show that the most unstable axial wavenumber is unbounded with increasing Reynolds number \cite{bil05}. Furthermore, an increase of the axial decorrelation for the azimuthal velocity length with increasing $Re_\tau$ in DNS of plane Couette flow was seen in Refs. \cite{avs14,pir14}. We can thus speculate that for infinite Reynolds number, the roll wavelength becomes infinitely large, and thus we recover the axial symmetry of the system in a statistical sense. 

\begin{figure}
  \centering
    \includegraphics[width=0.99\textwidth,trim={2cm 1cm 2cm 0},clip]{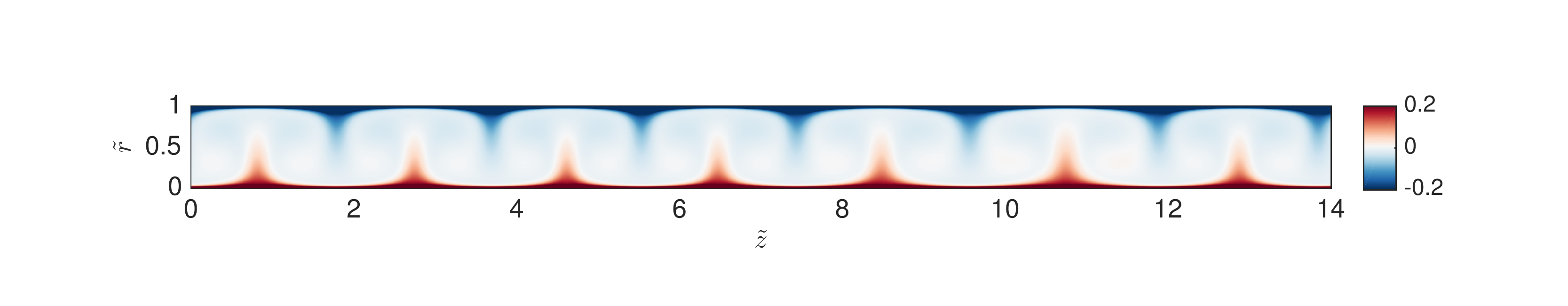}
  \caption{ Azimuthally- and temporally averaged azimuthal velocity for the G14R7 case. }
\label{fig:q1mehuge}
\end{figure}

\begin{figure}
  \centering
    \includegraphics[width=0.49\textwidth]{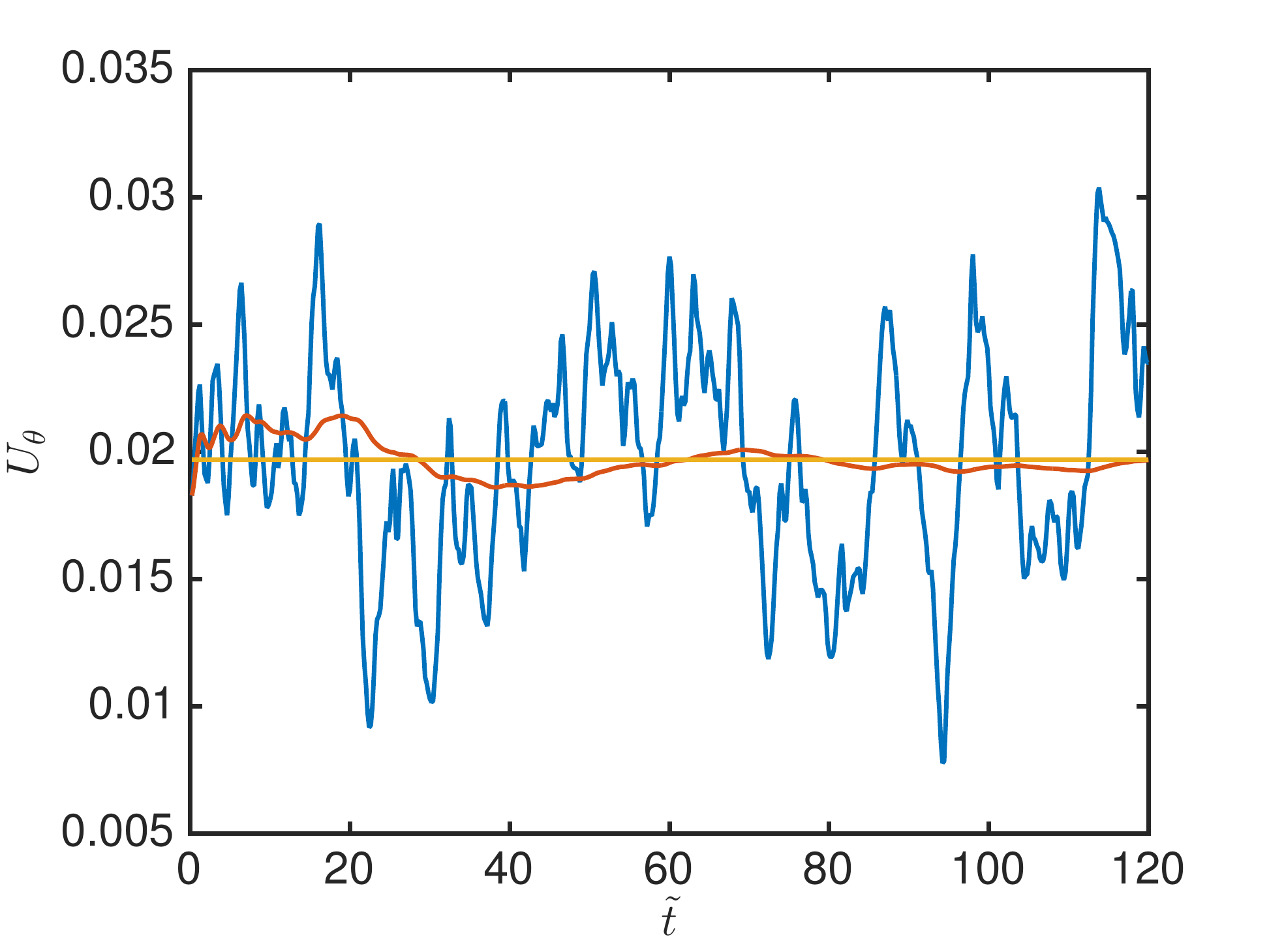}
  \caption{ The blue line shows the temporal evolution of the azimuthally-averaged azimuthal velocity at a point in the mid-gap inside the ejection region of a roll. The orangish line shows the running average of this quantity, and the ochre line shows the final average. The data used is for only $120$ time-units, as this is representative of all simulations in this manuscript. }
\label{fig:q1insthuge}
\end{figure}

To further demonstrate the axial pinning of the rolls, in Figure \ref{fig:q1mehuge} we show the azimuthally- and temporally averaged azimuthal velocity for the G17R7 case. Statistics for this case have been taken for a much longer time than the other four cases, and still the rolls, which have a square-aspect ratio here, appear to be fixed. The axial extent of this box, ($\approx 9\pi$ half-gaps) is larger than the one simulated in the largest simulations of Plane Couette flow by Komminaho \emph{et al.} \cite{kom96} ($8\pi$) and Tsukahara \emph{et al.} \cite{tsu06} ($\approx 8.2\pi$), and we expect it to be sufficient to show that the large-scale structures are indeed pinned. To ensure that we are not missing a slow evolution of the rolls, Figure \ref{fig:q1insthuge} shows the azimuthally-averaged velocity at a mid-gap point inside a roll, as well as the running mean, and the final value of the mean. Small-amplitude temporal fluctuations can be seen, with a characteristic timescale of $\sim 30$ non-dimensional time-units. This coincides with the frictional time-scale $\mathcal{O}(u_\tau/d)$. However, these oscillations are too small to be an unpinning of the rolls, which would amount for much larger variations of velocity. In addition, we note that we have only shown data for $120$ time-units, instead of the $487$ time-units that statistics were taken for the G14R7 case, to ensure fair comparison to the other cases in this manuscript. The very large time average window of the G14R7 case has resulted in a spatial convergence of the angular velocity current which is constant to less than $.5\%$, a more stringent requirement than other cases.   

\begin{figure}
  \centering
    \includegraphics[width=0.49\textwidth]{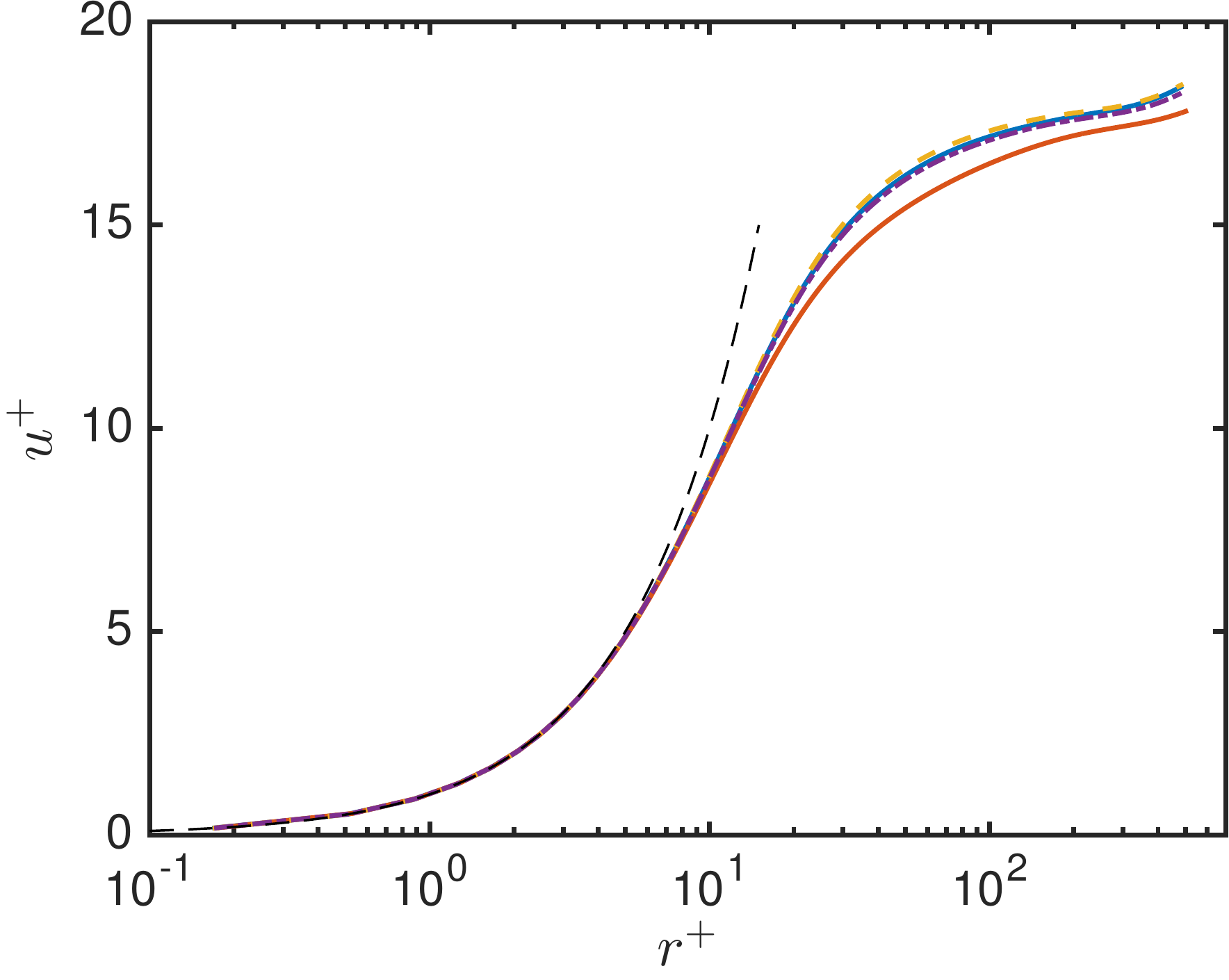}
  \caption{ Mean streamwise velocity profile at the inner cylinder in wall units  for the four simulated cases. Symbols as in Table \ref{tbl:reso}. Thin black dashed line indicates $r^+=u^+$. }
\label{fig:q1wall}
\end{figure}

Figure \ref{fig:q1wall} shows the average streamwise velocity profile at the inner cylinder in wall-units, i.e.~$u^+=(r_i\omega_i-\langle u_\theta \rangle_{\theta,z,t})/u_{\tau,i}$ against $r^+=(r-r_i)/\delta_{\nu,i}$. We can see a clear dependence on $\lambda_z$ and on the roll wavelength, but again not on the number of rolls. This dependence on $\lambda_z$ becomes weaker for increasing $Re_\tau$, as it has almost vanished in the $Re_\tau\approx1400$ cases of Ref.~\cite{ost15}. This is further confirmation that small boxes with a single roll produce accurate statistics for the mean velocity profiles, and that a single roll pair is enough.

\begin{figure}
  \centering
    \includegraphics[width=0.32\textwidth]{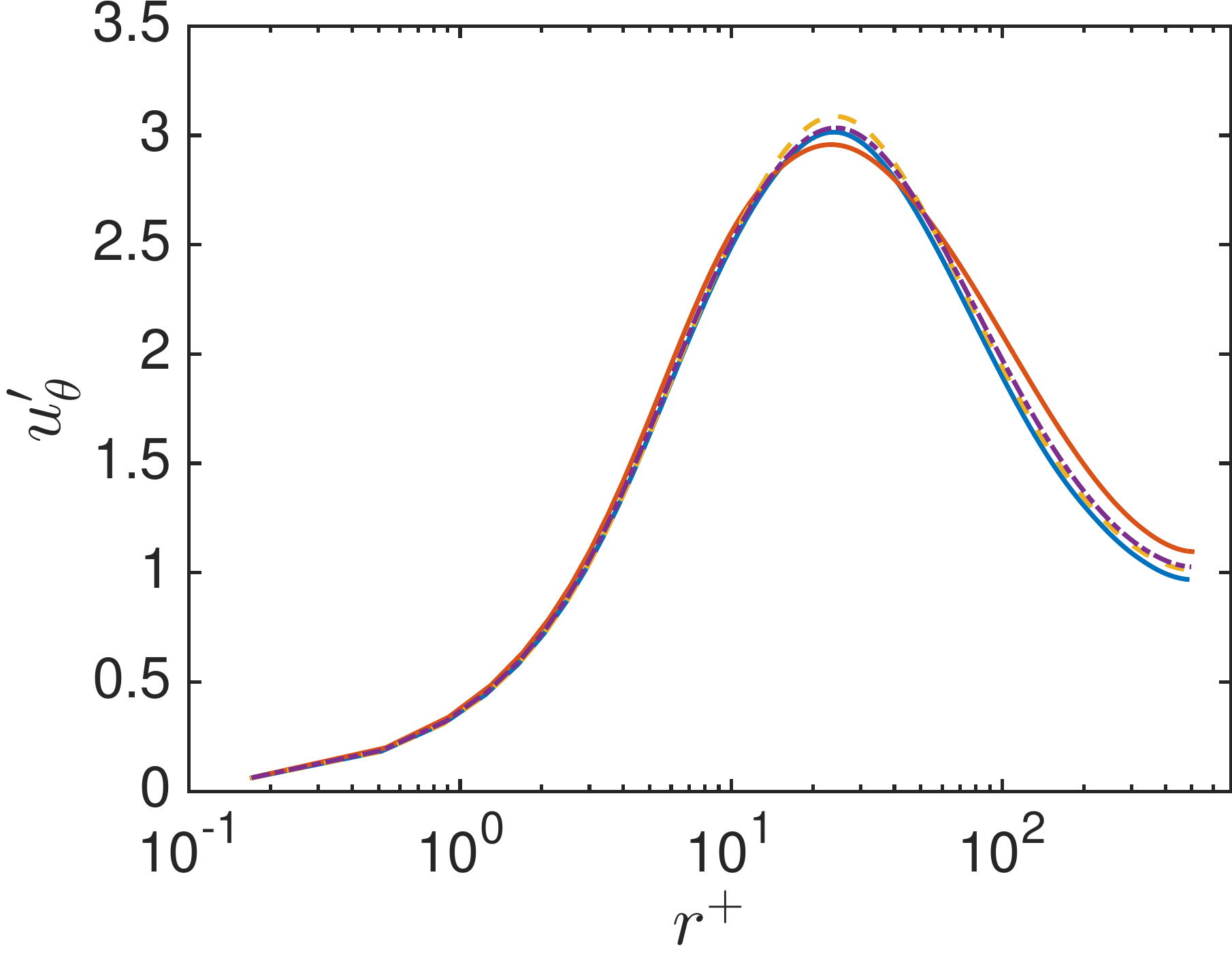}
     \includegraphics[width=0.32\textwidth]{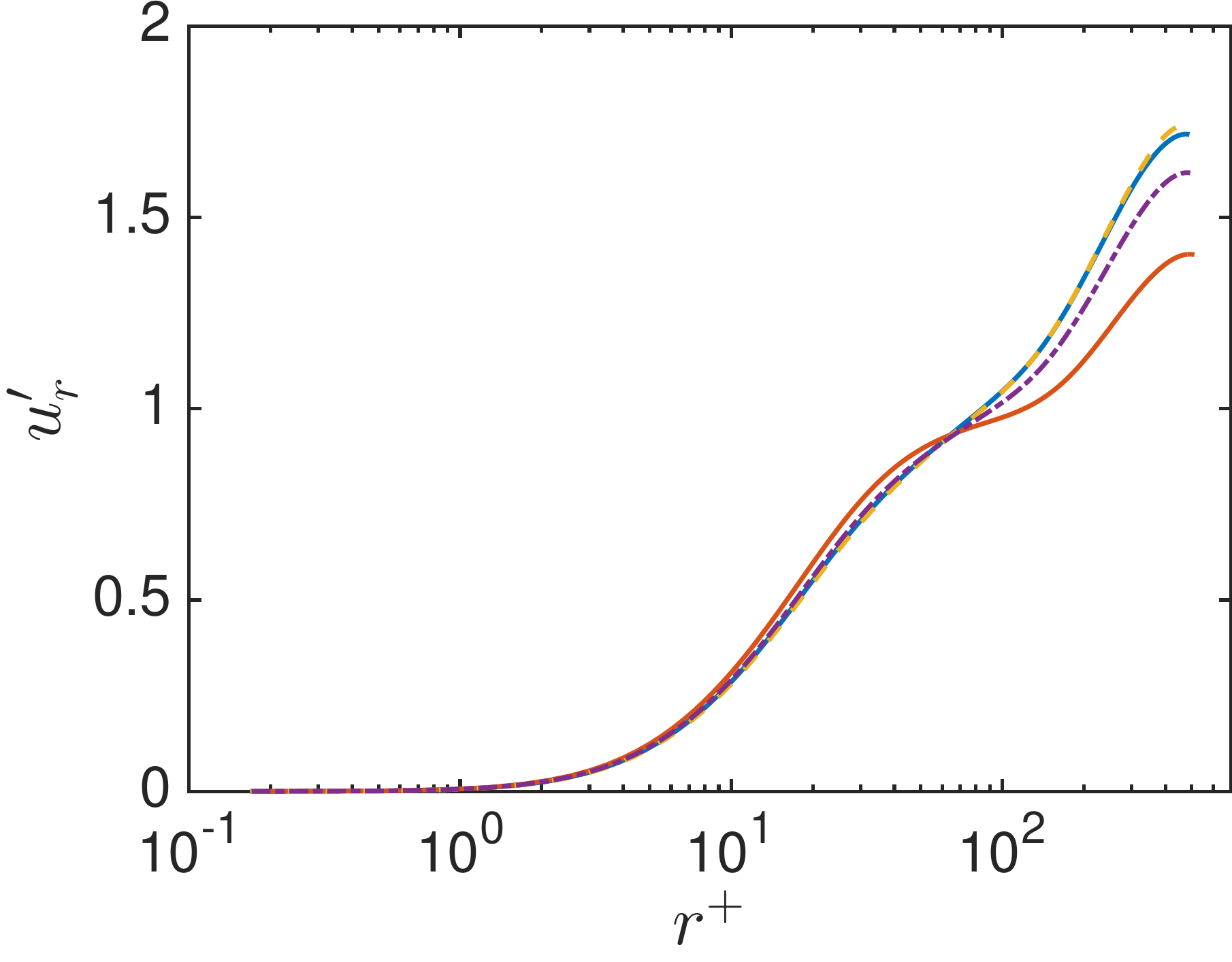}
     \includegraphics[width=0.32\textwidth]{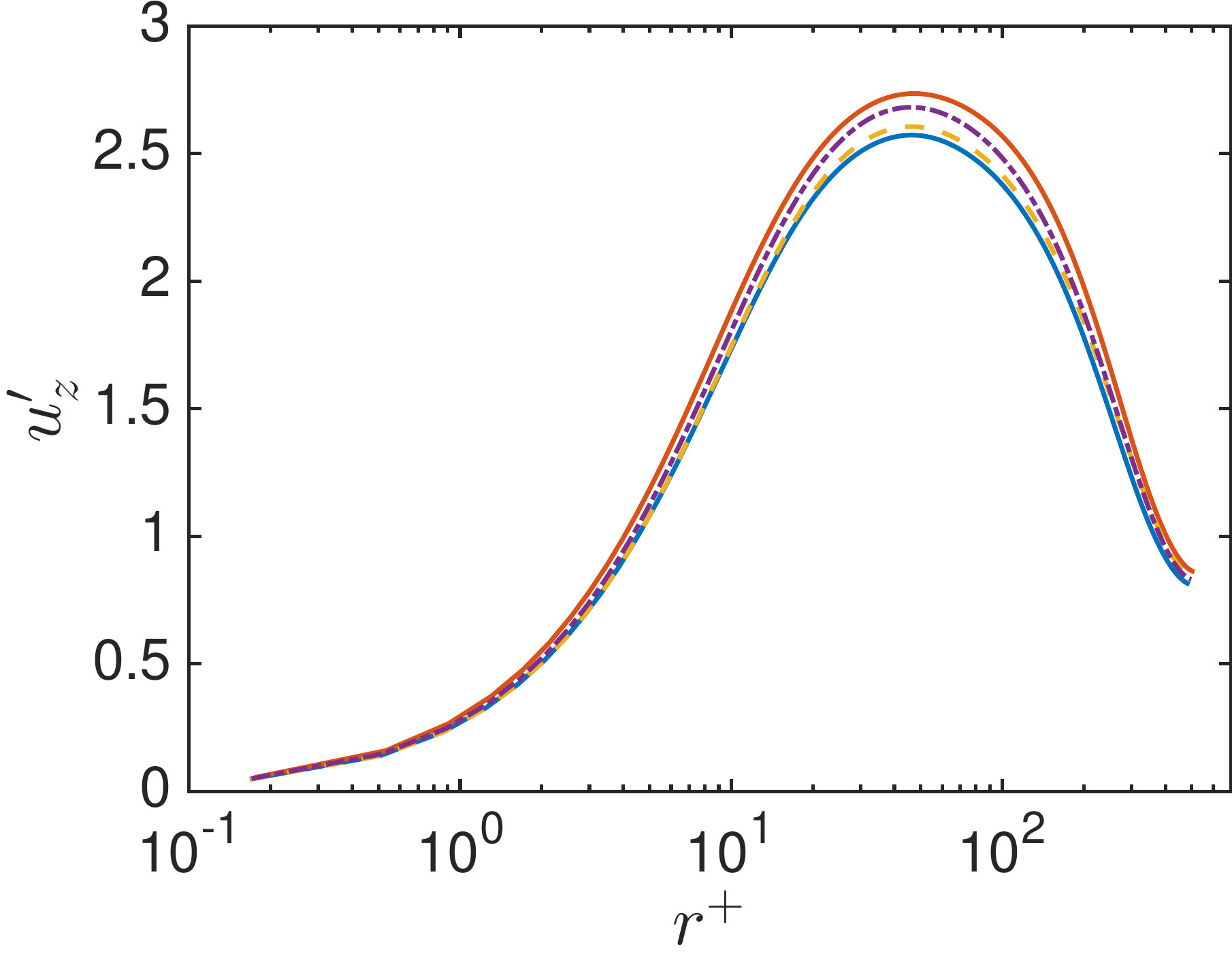}
  \caption{ Fluctuation velocity profiles at the inner cylinder in wall units for the four simulated cases. Symbols as in Table \ref{tbl:reso}.  }
\label{fig:qprimewall}
\end{figure}

In contrast, figure \ref{fig:qprimewall} shows the fluctuations of the streamwise velocity profile at the inner cylinder in wall-units, i.e.~$u^\prime=(\langle u^2 \rangle_{\theta,z,t}-\langle u\rangle^2_{\theta,z,t})^{1/2}/u_{\tau,i}$. The fluctuations show not only a dependence on $\lambda_z$, but also on the amount of rolls, increasing for the three-roll pair case when compared to the single-roll pair case with the same $\lambda_z$. These might be due to interactions between roll pairs and the different strength of each of the roll pairs. This is again consistent with the findings of Ref.~\cite{ost15}, where the fluctuation statistics had not saturated to box-independent values for $\Gamma=4$. Some degree of collapse is seen near the inner cylinder ($r^+<60$). However, the fluctuation peak of $u^\prime_\theta$ at $r^+\approx 12$ is substantially different across the considered cases, indicating that the fluctuations are both produced by the rolls itself, and by roll-to-roll interactions.

\begin{figure}
  \centering
    \includegraphics[width=0.32\textwidth]{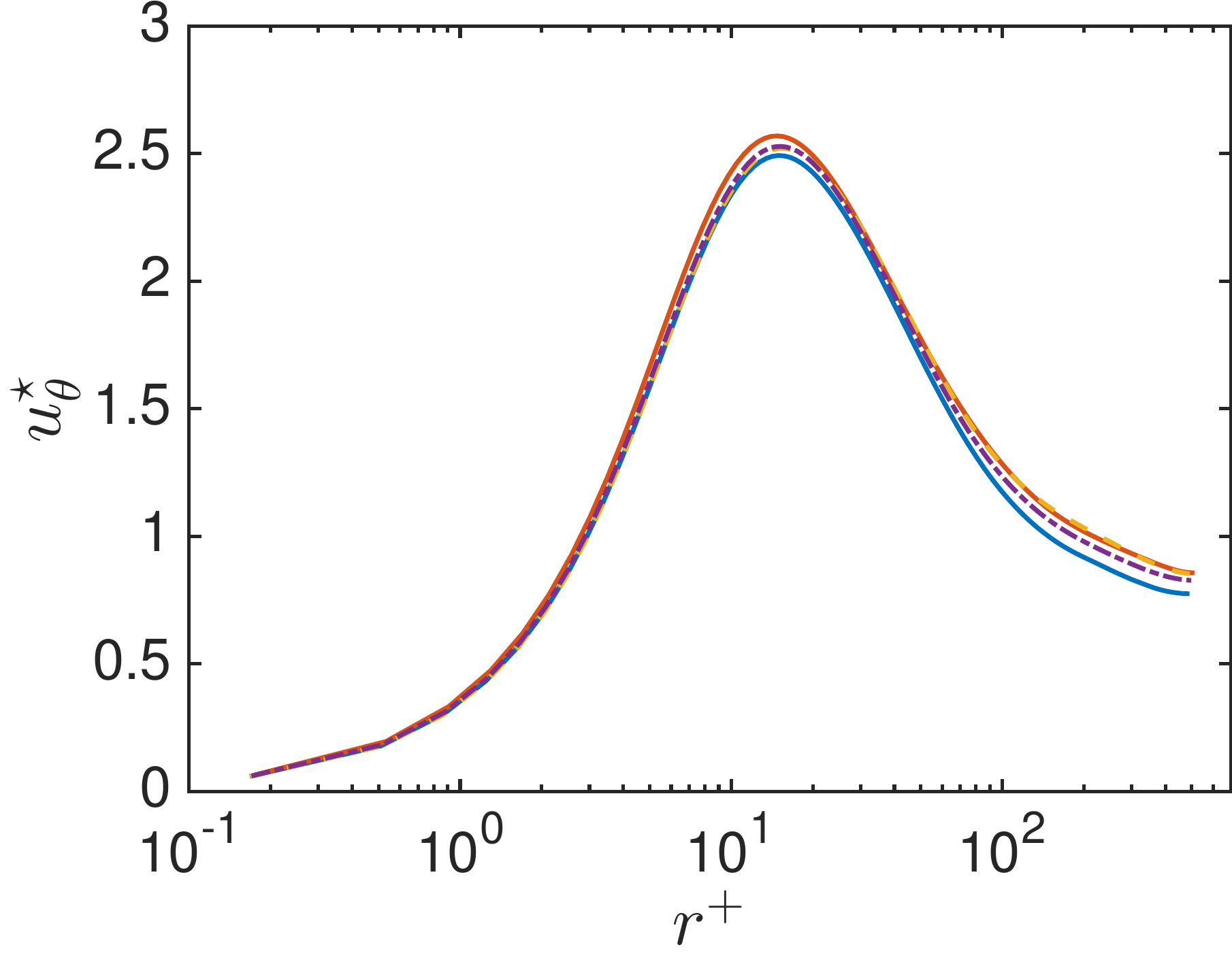}
         \includegraphics[width=0.32\textwidth]{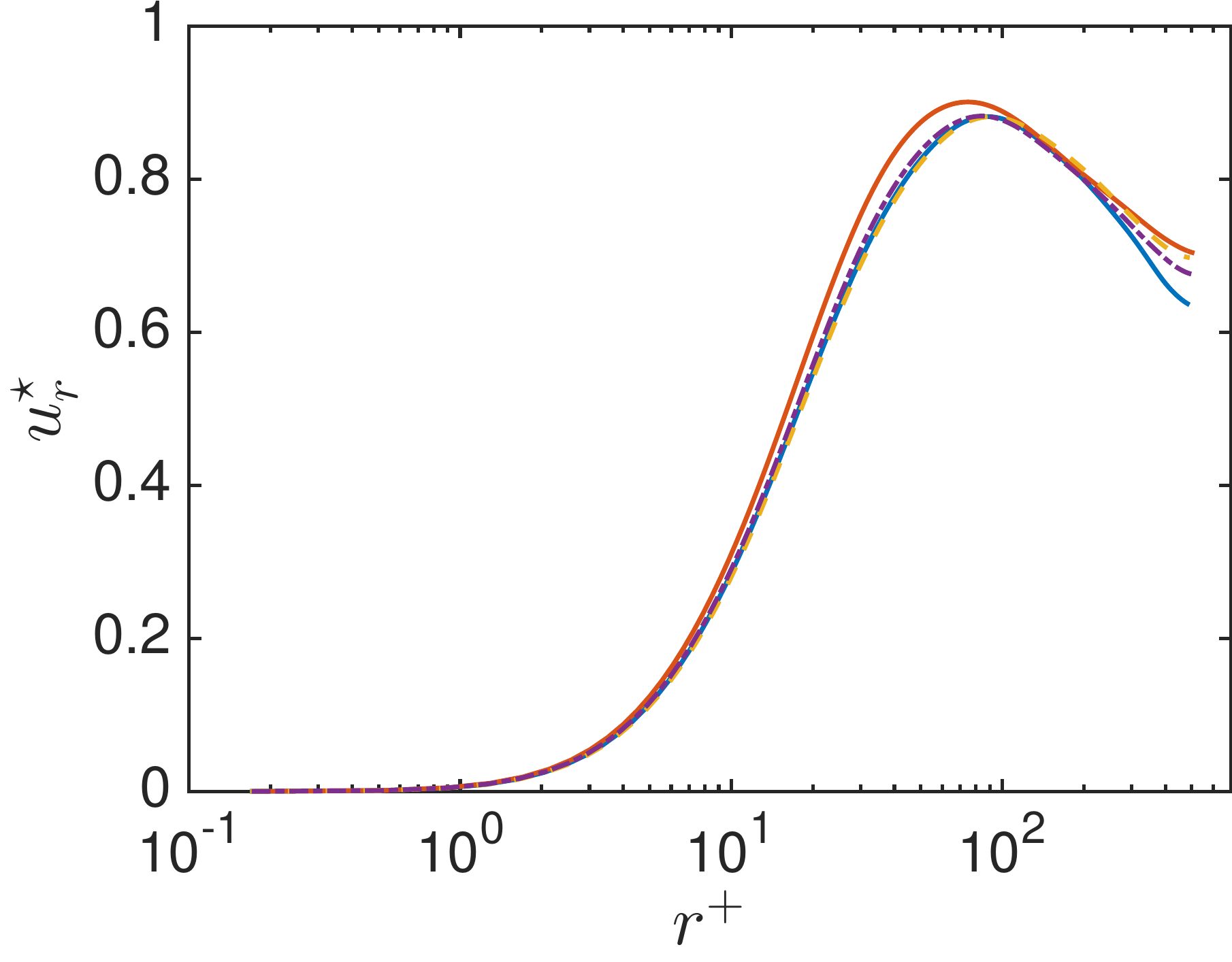}
     \includegraphics[width=0.32\textwidth]{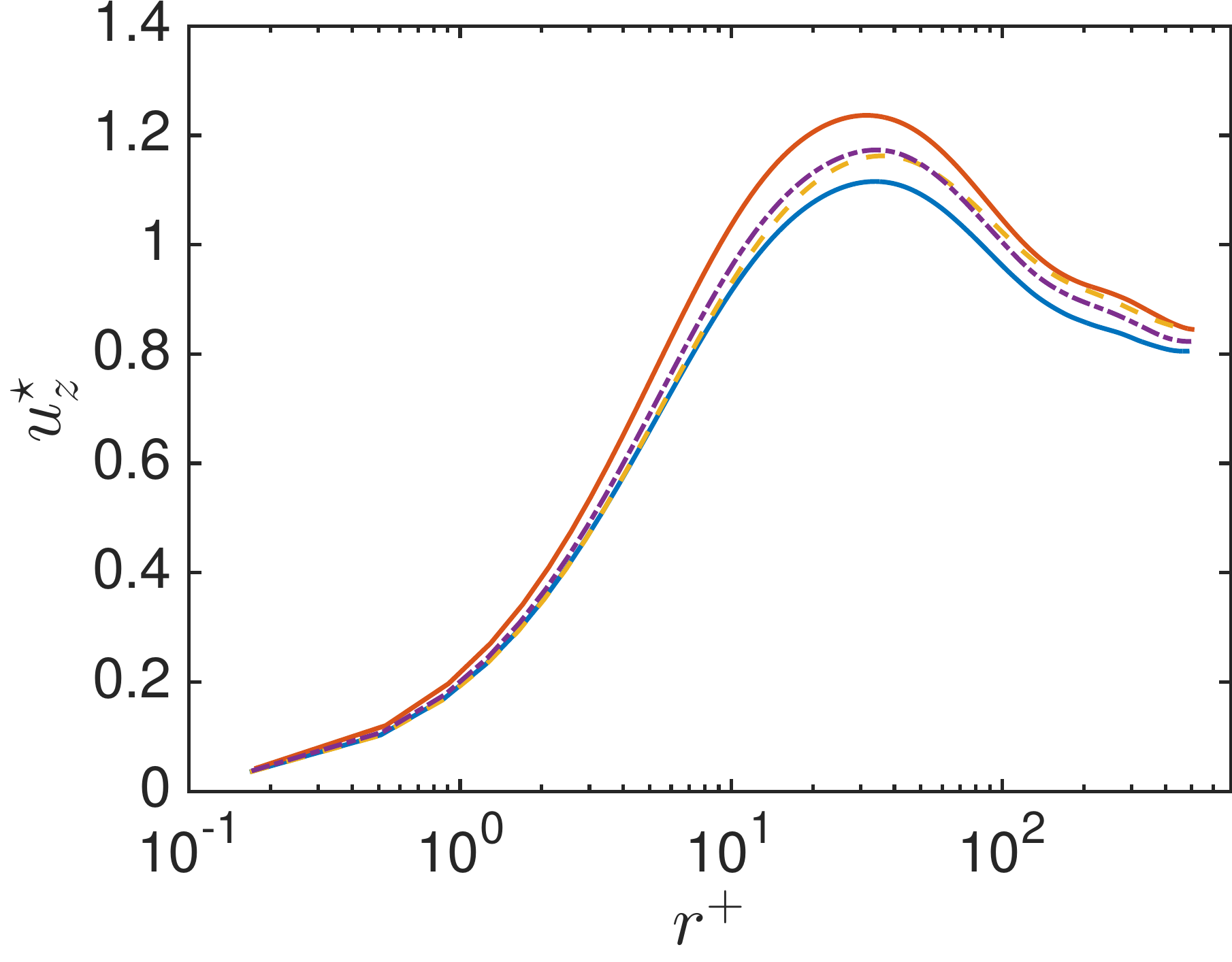}
  \caption{  Roll-less fluctuation velocity profiles at the inner cylinder in wall units for the four simulated cases.  Symbols as in Table \ref{tbl:reso}.   }
\label{fig:qstarwall}
\end{figure}

To remove the effect of the large-scale structures on the fluctuations, we define the roll-less azimuthal velocity fluctuations as $u^\star=\langle \langle u^2 \rangle_{\theta,t}-\langle u \rangle^2_{\theta,t}\rangle_z^{1/2}/u_{\tau,i}$ \cite{ost16}. As detailed in Ref.~\cite{ost16}, both definitions for the fluctuations give equivalent results for statistically homogeneous flows, but give different results for flows which show some axial inhomogeneity. Figure \ref{fig:qstarwall} shows $u^\star$ at the inner cylinder for the four simulated cases. A similar behaviour is seen across the three velocity components. Near the inner cylinder ($r^+<60$), (more markedly for $u^\star_z$), the profiles depend mainly on $\lambda_z$. The curve which shows the largest deviation from the others corresponds to the G7R2 case. However, for increasing $r^+$, i.e.\ as the bulk is approached, the collapse is better for the large $\Gamma$ cases, regardless of $\lambda_z$. From this it seems that the nature of the fluctuations captured by $u^\star$ changes in the bulk. Near the cylinder, it seems to capture the same sort of fluctuations as $u^\prime$, as this near-wall region is more axially homogeneous. Meanwhile, in the bulk it may be capturing fluctuations caused by shear and not by centrifugal forces. This is especially pronounced in the case of $u_\theta$ fluctuations. While in Fig. \ref{fig:qprimewall}, some degree of collapse could be seen for the two cases with $\lambda_z=2.33$ and one or three roll-pairs, we see that the collapse in $u_\theta^\star$ happens for cases with similar $\Gamma$ and different $\lambda_z$.

\begin{figure}
  \centering
    \includegraphics[width=0.4\textwidth]{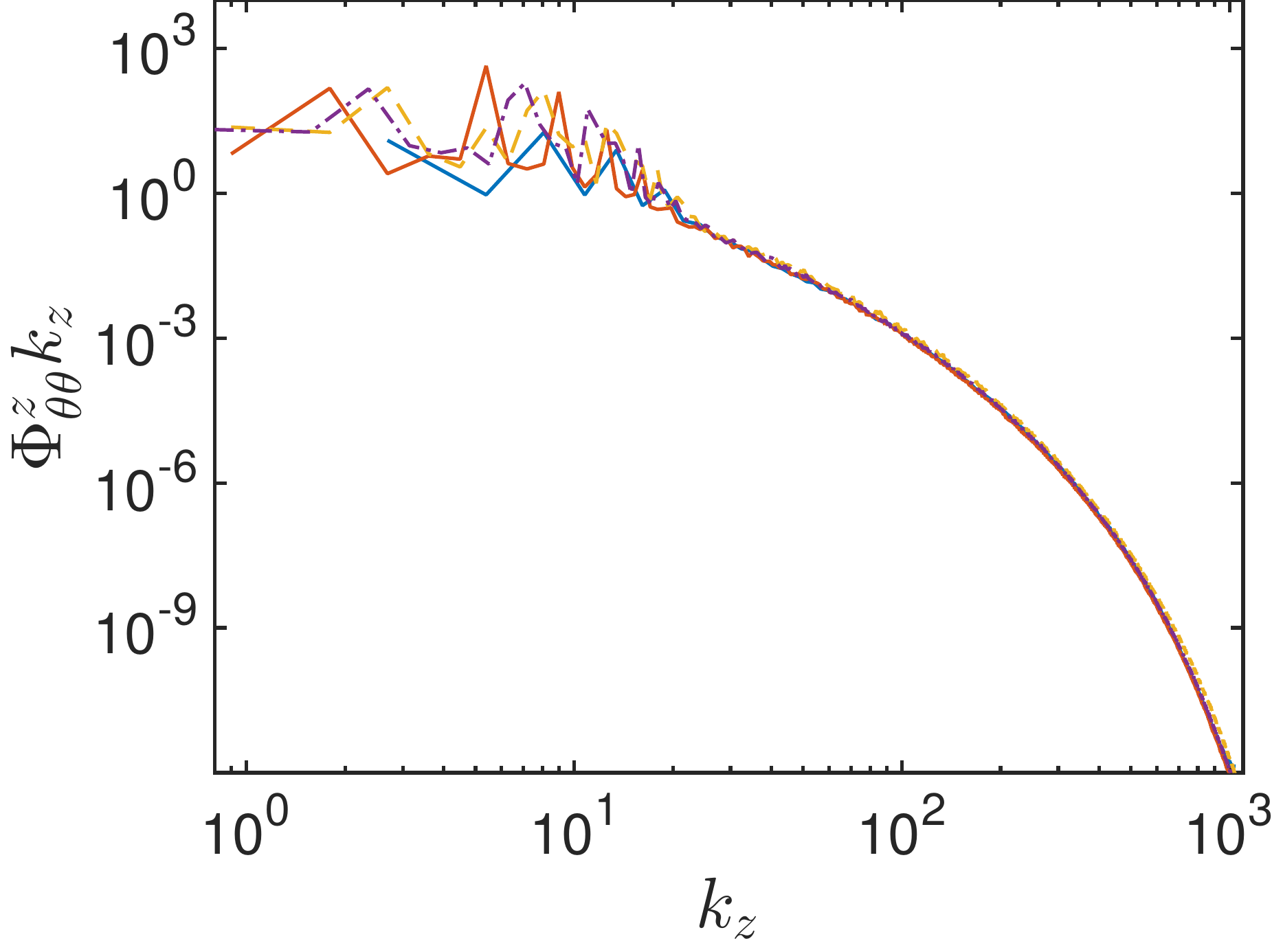}
    \includegraphics[width=0.4\textwidth]{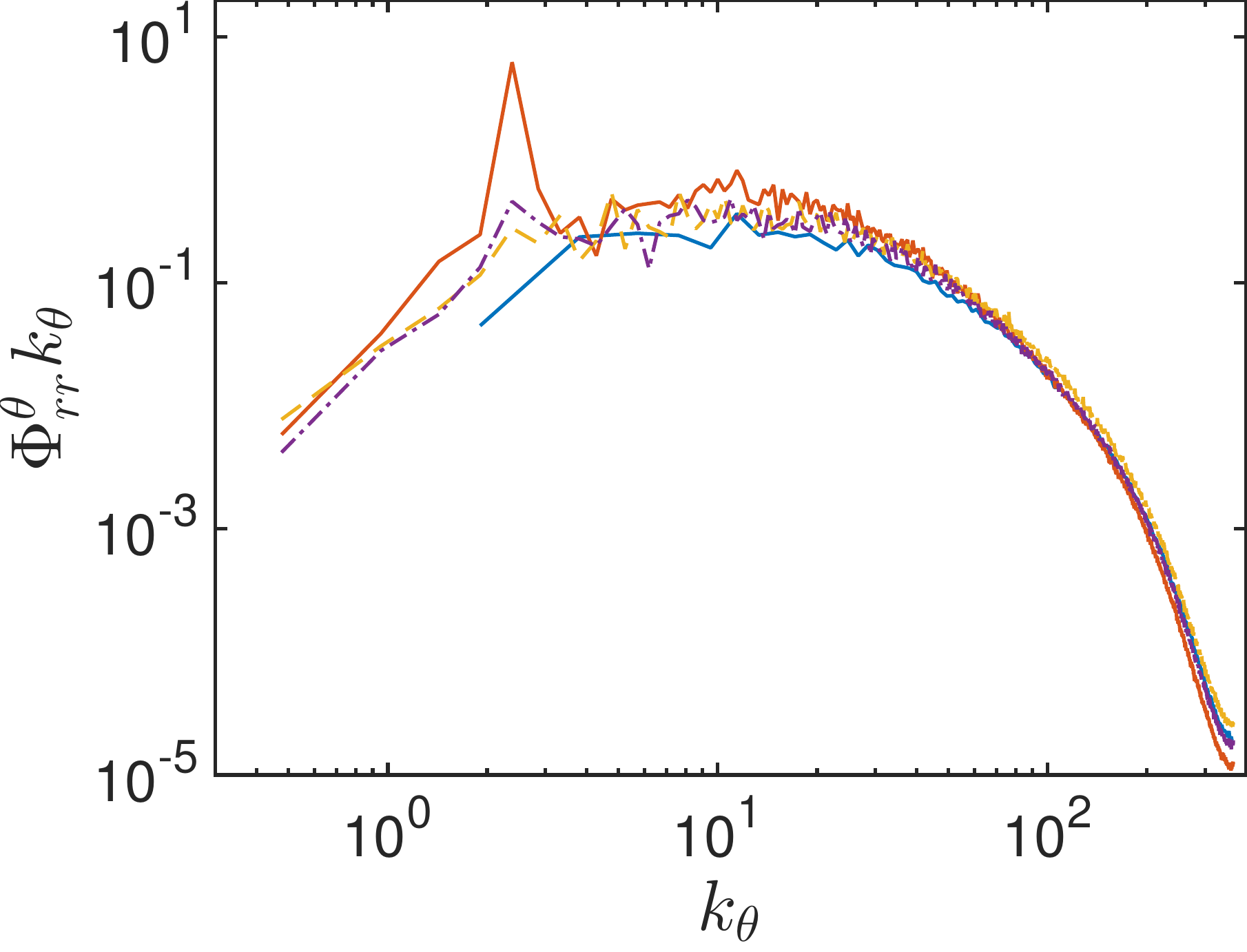}\\
     \includegraphics[width=0.4\textwidth]{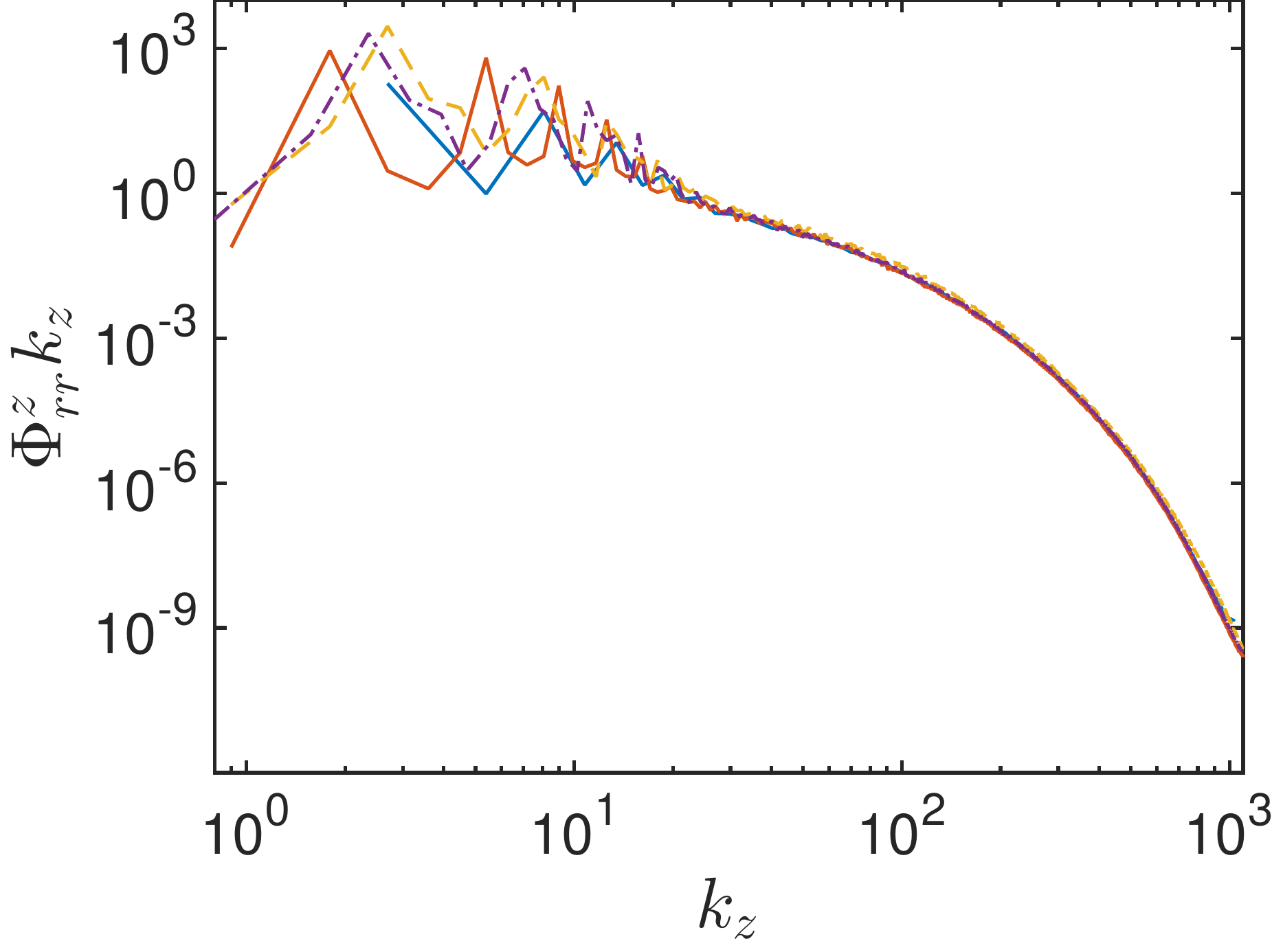}
      \includegraphics[width=0.4\textwidth]{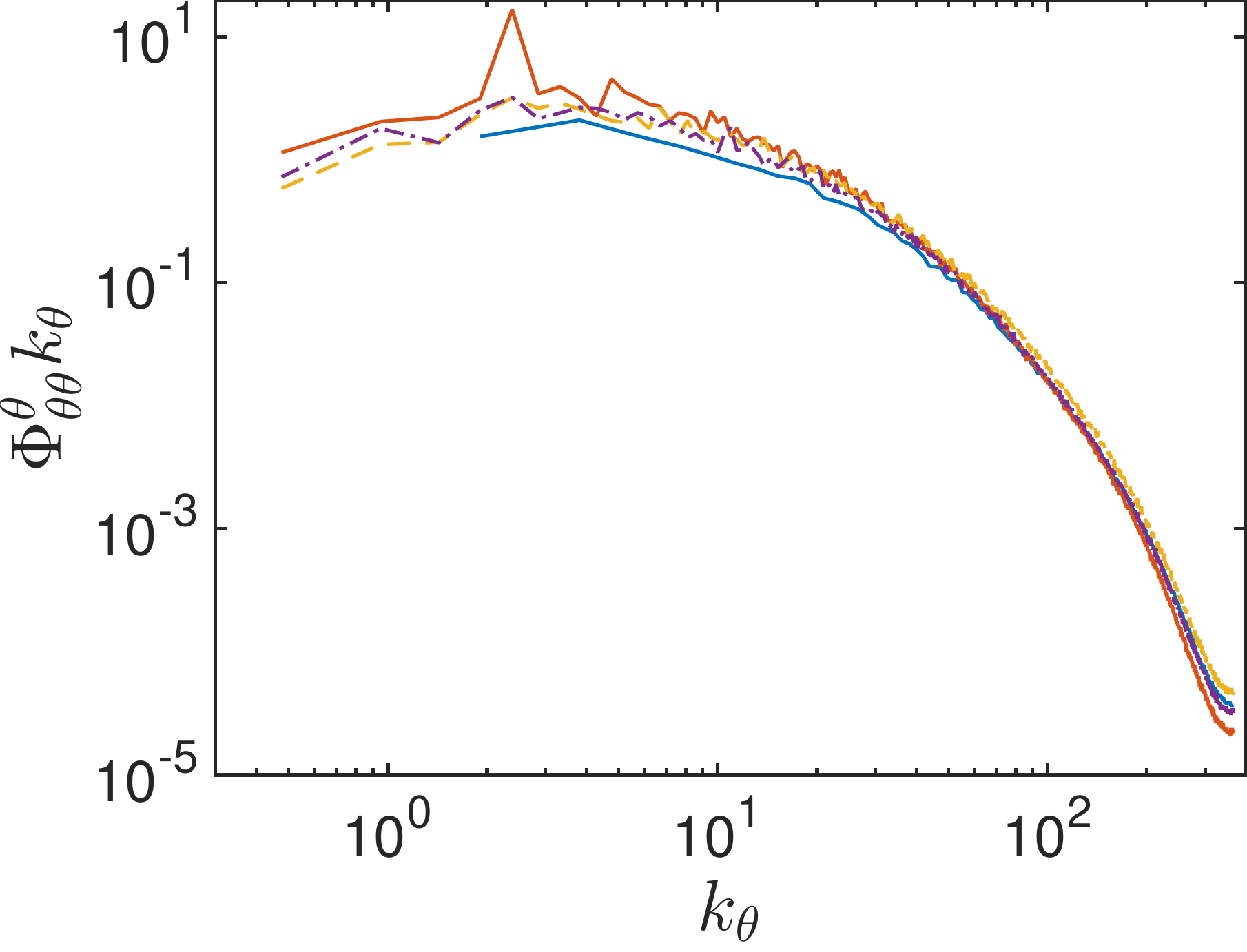}
  \caption{ Premultiplied velocity energy spectra $\Phi k$ in the axial (left) and azimuthal (right) directions for the azimuthal (left) and radial (right) velocities.  Symbols as in Table \ref{tbl:reso}.   }
\label{fig:spectraslab9}
\end{figure}

Finally, we show the premultiplied azimuthal and axial energy spectra of the azimuthal and axial velocity at mid-gap in Fig. \ref{fig:spectraslab9}. For the axial spectra, the low wavenumber part is dominated by the Taylor rolls. A clear maximum at the roll wave number $k_{TR}=2\pi/\lambda_z$, followed by a drop-off for smaller wavenumbers. The highest energy modes correspond to the harmonics of associated to the roll, i.e. $nk_{TR}$ with $n\in\mathbb{Z}$, and this results in a clear sawtooth pattern. The high wavenumber spectra collapse across all cases, consistent with the results of Ref.~\cite{ost15}. This leads to the $u^\star$ fluctuations for $\Gamma=7$. Remarkably, G2R1 and G7R3, i.e.~the two cases with the same $\lambda_z$ and different roll number do not collapse in this region. It appears that the details of the Taylor roll play a very small role in the fine features of the small scale fluctuations, and that a single roll system cannot adequately capture the energy spectra of multiple rolls. This supports the conclusion that roll to roll interaction is important for determining the fluctuations, and especially the fluctuation peak at $r^+=12$.

The azimuthal spectra for both velocities are consistent with what was seen in Refs.~\cite{ost15,ost16}. The premultiplied spectra, (i.e.~multiplied by $k$) considered here have a clear maximum. We are thus simulating the largest energy-containing scales by using $n_{sym}=5$. However, there is a marked peak for the G7R2 case for wavelengths corresponding to one-quarter of the azimuthal extent of the domain, i.e. modes with a rotational symmetry of order $20$. This has not been reported before, and might be due to a resonance of the box- a similar phenomena was seen in Ref.~\cite{ost15} for a computational box with $\Gamma=2.09$ and $n_{sym}=10$, which saw a very large increase of the velocity fluctuations at the mid-gap. These peaks could indicate certain ``unnatural'' resonances of the computational box, which exactly fit ``wavy'' azimuthal patterns.

In summary, we have run a series of simulations of TC flow using computational boxes with a large axial extent. The large-scale Taylor rolls were found to still be fixed, even for the largest axial aspect ratio of $\Gamma=8$. The rolls were found to preferably be in a ``tall'' configuration, i.e., to have a preferred wavelength of $\lambda_z>2$. Furthermore, two possible configurations were found to be stable for long periods of time for $\Gamma=7$, a two-roll pair configuration with $\lambda_z=3.5$ and a three-roll pair configuration with $\lambda_z=2.33$. A single roll pair was found to give the same torque and mean velocity profiles as a simulation of three roll pairs at with the same $\lambda_z$, consistent with the findings of Ref.~\cite{bra13}. A weak dependence on $\lambda_z$ was found for the torque, despite different fluctuation profiles in the bulk, supporting that the torque is dominated by the near-wall region. We have found that a small box with a single roll pair is large enough to reproduce the torque and mean velocity profile turbulent TC flow, but it cannot reproduce the fluctuations and velocity spectra. This study in combination with Ref.~\cite{ost15} provides some promising evidence that at sufficiently high Reynolds number, the statistics of Taylor-Couette flow can reach box-size and roll-wavelength independence, provided the axial extent of the boxes is large enough. This will only be confirmed once such these high Reynolds number simulations are actually conducted.

\section*{Acknowledgements}

We thank V. Spandan for extensive help in proof-reading the paper. We acknowledge M. Bernardini, M. P. Encinar, J. Jim\'enez, P. Orlandi, S. Pirozzoli, Y. Yang, and X. Zhu for fruitful and stimulating discussions. We also gratefully acknowledge computational time for the simulations provided by SurfSARA on resource Cartesius through a NWO grant. 

\bibliography{Literatur}

\end{document}